\documentclass{article}
\usepackage{spconf,amsmath,epsfig,indentfirst,adjustbox,url,hyperref}

\title{RANDOMIZED PRINCIPAL COMPONENT ANALYSIS FOR HYPERSPECTRAL IMAGE CLASSIFICATION}
\name{Mustafa Ustuner}
\address{Department of Geomatics Engineering, Artvin Coruh University, 08100 Artvin, Turkey}
\begin{document}
%\ninept
%
\maketitle
\begin{abstract}
The high-dimensional feature space of the hyperspectral imagery poses major challenges to the processing and analysis of the hyperspectral data sets. In such a case, dimensionality reduction is necessary to decrease the computational complexity. The random projections open up new ways of dimensionality reduction, especially for large data sets. In this paper, the principal component analysis (PCA) and randomized principal component analysis (R-PCA) for the classification of hyperspectral images using support vector machines (SVM) and light gradient boosting machines (LightGBM) have been investigated. In this experimental research, the number of features was reduced to 20 and 30 for classification of two hyperspectral datasets (Indian Pines and Pavia University). The experimental results demonstrated that PCA outperformed R-PCA for SVM for both datasets, but received close accuracy values for LightGBM. The highest classification accuracies were obtained as 0.9925 and 0.9639 by LightGBM with original features for the Pavia University and Indian Pines, respectively.
\end{abstract}
\begin{keywords}
Hyperspectral, PCA, R-PCA, LightGBM, SVM
\end{keywords}
\section{Introduction}
\label{sec:intro}

In recent years, there has been a notable increase in the use of hyperspectral remote sensing images, acquired from airborne or spaceborne sensors, with rapid development in the Earth observation and imaging technology \cite{jia2013feature,linhe2018,rasti2020feature}. Hyperspectral images provide rich information about the land objects through narrow spectral bands \cite{jia2013feature,uddin2021}. Though this increased dimensionality of data provides detailed information, it poses challenges (curse of dimensionality) to conventional techniques, such as classification, for the accurate analysis of hyperspectral images \cite{jia2013feature,rasti2020feature,ghamisi2017advances}. \par 
Hyperspectral images are used in different kinds of applications such as environmental monitoring, crop classification, and target detection. Though hyperspectral images provide essential and rich information regarding the land objects, there are some noisy and redundant spectral bands among the hundreds of bands that might pose challenges in their classification \cite{jia2013feature,hidalgo2021dimensionality,camps1}. Applying dimension (feature) reduction techniques is necessary to remove these redundancies in the hyperspectral data. The main objective of dimension (feature) reduction is to identify the best subset of original or transformed features while retaining as many spectral details as possible \cite{jia2013feature,ghamisi2017advances,hidalgo2021dimensionality}. Dimensionality reduction of the hyperspectral data can be implemented in two ways: (1) feature selection or (2) feature extraction. Feature selection aims to identify the most suitable subset of original features based on selection criteria, while feature extraction transforms the data into a much lower-dimensional subspace. Dimension reduction techniques can be classified into three categories: unsupervised, supervised, and semi-supervised approaches \cite{ghamisi2017advances}.\par
Unsupervised dimensionality reduction techniques can be further grouped into four categories, based on the operation principle: (1) based on information content, (2) projection based, (3) similarity measures and (4) frequency analysis \cite{hidalgo2021dimensionality}. Additionally, numerous methods, such as PCA, independent component analysis (ICA), kernel PCA, linear discriminant analysis (LDA), R-PCA and local linear feature extraction, have been employed for reducing the dimensionality of hyperspectral images \cite{jia2013feature,ghamisi2017advances}. PCA, a projection-based technique, is the most frequently utilized in hyperspectral image analysis among these methods. It aims to produce a smaller set of orthogonal variables in lower-dimensional subspace from the original features with maximum variance \cite{Senthilnath,rasti2020feature}. However, the implementation of PCA can be computationally intensive and time-consuming. Therefore, the use of random projections is considered as an alternative approach for dimensionality reduction in terms of computation time. Random projections are computationally cheap and require less memory compared to classical dimensionality reduction techniques \cite{halko,Wojnowicz}. The randomized PCA, which is a random projection of PCA, is computed via randomized (approximated) singular value decomposition (SVD) of the original data \cite{halko,Wojnowicz, firat2022}. \par
Several studies have examined the potential for use of R-PCA for feature reduction and classification of the hyperspectral images. Makantasis et al. introduced the R-PCA for dimensionality reduction and combined it with a convolutional neural network (CNN) for encoding spatial and spectral information (denoted as R-PCA CNN) \cite{makantasis}. Most recent studies that utilize R-PCA explore deep learning models for the classification of hyperspectral images. Mei et al. proposed a five-layer CNN (C-CNN) and feature-learning CNN (FL-CNN) for hyperspectral image classification and compared the performances with state-of-the-art CNN-based methods, R-PCA CNN and CNN based on pixel-pair features (CNN-PPF) \cite{mei2017}. Zhang et al. introduce a framework, called diverse region-based deep CNN model (denoted as DR-CNN) for the classification of hyperspectral images and compare the performance of the DR-CNN with other deep learning and state-of-the-art classifiers such as SVM, CNN, R-PCA CNN, CNN-PPF and SS-CNN \cite{zhang2018}. Liu et al. proposed a new method called deep few-shot learning method to address the small sample size problem of hyperspectral image classification. In their experimental research, they compared the performance of their proposed method with several CNN-based methods including R-PCA CNN \cite{liu2019}. Only a few papers have investigated the R-PCA with conventional machine learning algorithms, excluding deep learning. In \cite{makantasis}, the classification performance of R-PCA CNN was compared with SVM-based methods (RBF-SVM and Linear-SVM). Damodaran et al. compared the R-PCA and kernel PCA for the dimensionality reduction of hyperspectral images using SVM classification \cite{Damodaran}. So, the impacts of R-PCA features on hyperspectral image classification accuracy on supervised machine learning algorithms have not been fully explored yet. \par

Thus, the main objective of this research is to investigate the impacts of R-PCA on hyperspectral image classification accuracy and compare it with PCA on two public available hyperspectral scenes (Indian Pines (Fig 1) and Pavia University (Fig 2)) \cite{hyperspectral} using machine learning algorithms.

\section{DATASET}
\label{sec:format}

In this experimental research, two publicly available and benchmark hyperspectral datasets (Indian Pines and Pavia University) were used and freely downloaded from \cite{hyperspectral} . 

\begin{figure}[htb]

\begin{minipage}[b]{1.0\linewidth}
\centering
 \includegraphics[width=\textwidth]{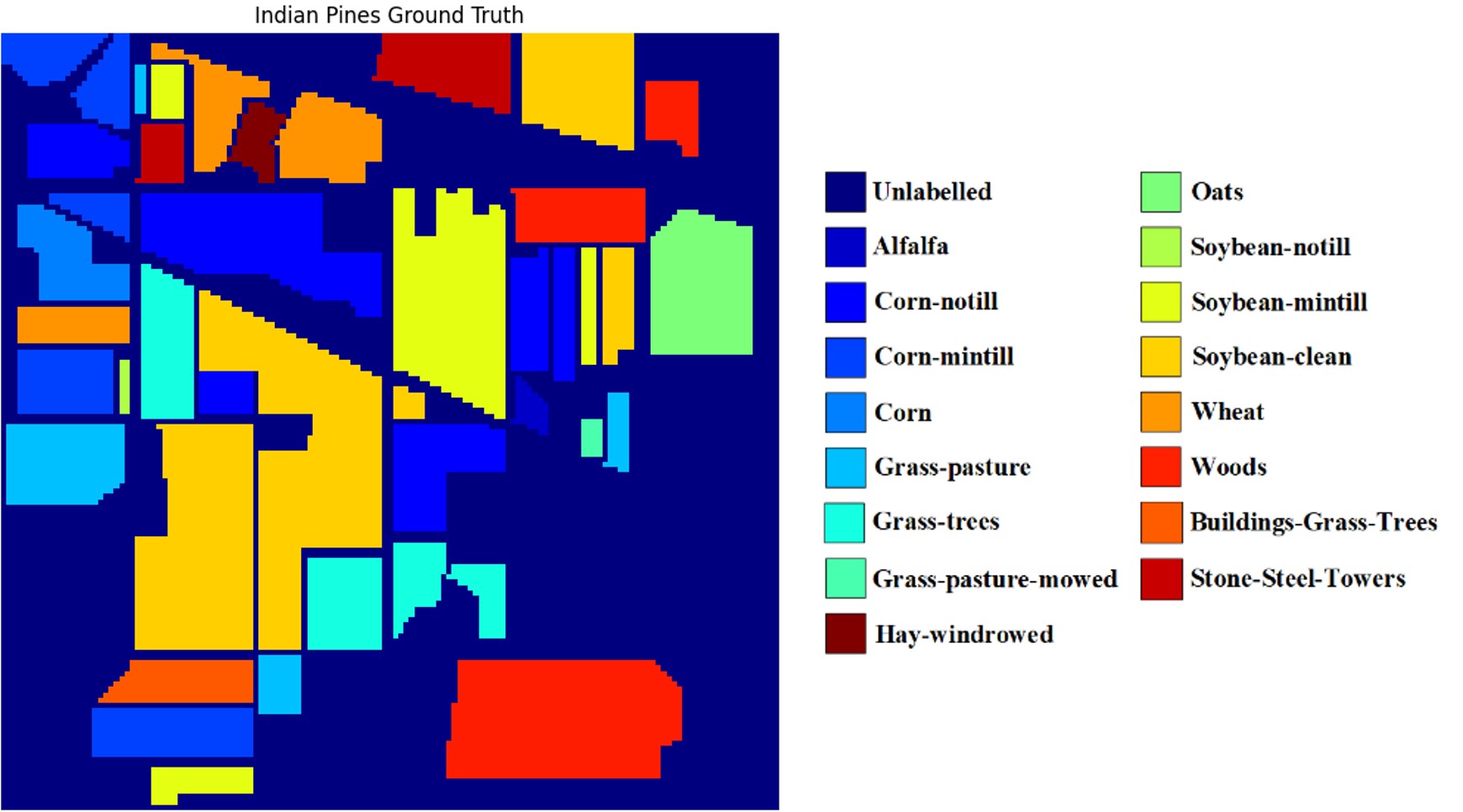}
 \vspace{-8mm}
 \caption{Ground truth image for Indian Pines scene}
  %\vspace{2.0cm}
  %\centerline{(a) Result 1}\medskip
\end{minipage}
\end{figure}

The spatial resolutions are 20 m and 1.3 m for Indian Pines and Pavia University scene, respectively. The Indian Pines dataset has originally 220 spectral bands from AVIRIS sensor however 20 spectral bands were removed to exclude the water absorption region \cite{hyperspectral,indianpines}. Therefore the remaining 200 spectral bands were used in this experimental research. \newline There are 16 ground-truth classes in Indian Pines dataset (Table 1). The landscape of the Indian Pines scene comprises mostly agricultural land, approximately two-thirds, with the remaining one-third consisting of forest or other types of perennial natural vegetation \cite{hyperspectral,indianpines}. 

\begin{table}[!ht]
 \vspace{-5mm}
\caption{Ground truth classes for the Indian Pines scene}
 \vspace{3mm}
\centering
\small
\begin{tabular}{|l|l|l|}
\hline
   & Class                        & Samples \\ \hline
1  & Alfalfa                      & 46      \\ \hline
2  & Corn-notill                  & 1428    \\ \hline
3  & Corn-mintill                 & 830     \\ \hline
4  & Corn                         & 237     \\ \hline
5  & Grass-pasture                & 483     \\ \hline
6  & Grass-trees                  & 730     \\ \hline
7  & Grass-pasture-mowed          & 28      \\ \hline
8  & Hay-windrowed                & 478     \\ \hline
9  & Oats                         & 20      \\ \hline
10 & Soybean-notill               & 972     \\ \hline
11 & Soybean-mintill              & 2455    \\ \hline
12 & Soybean-clean                & 593     \\ \hline
13 & Wheat                        & 205     \\ \hline
14 & Woods                        & 1265    \\ \hline
15 & Buildings-Grass-Trees-Drives & 386     \\ \hline
16 & Stone-Steel-Towers           & 93      \\ \hline
\end{tabular}
\end{table}

The Pavia University dataset contains 103 bands acquired from the airborne sensor ROSIS-03 (Reflective Optics Systems Imaging Spectrometer) and has nine ground-truth classes (Table 2). There are originally 115 spectral bands obtained from the sensor however 12 bands were removed due to noise \cite{hyperspectral,paviau1}.

\begin{figure}[!htb]

\begin{minipage}[b]{1.0\linewidth}
\centering
 \includegraphics[width=\textwidth]{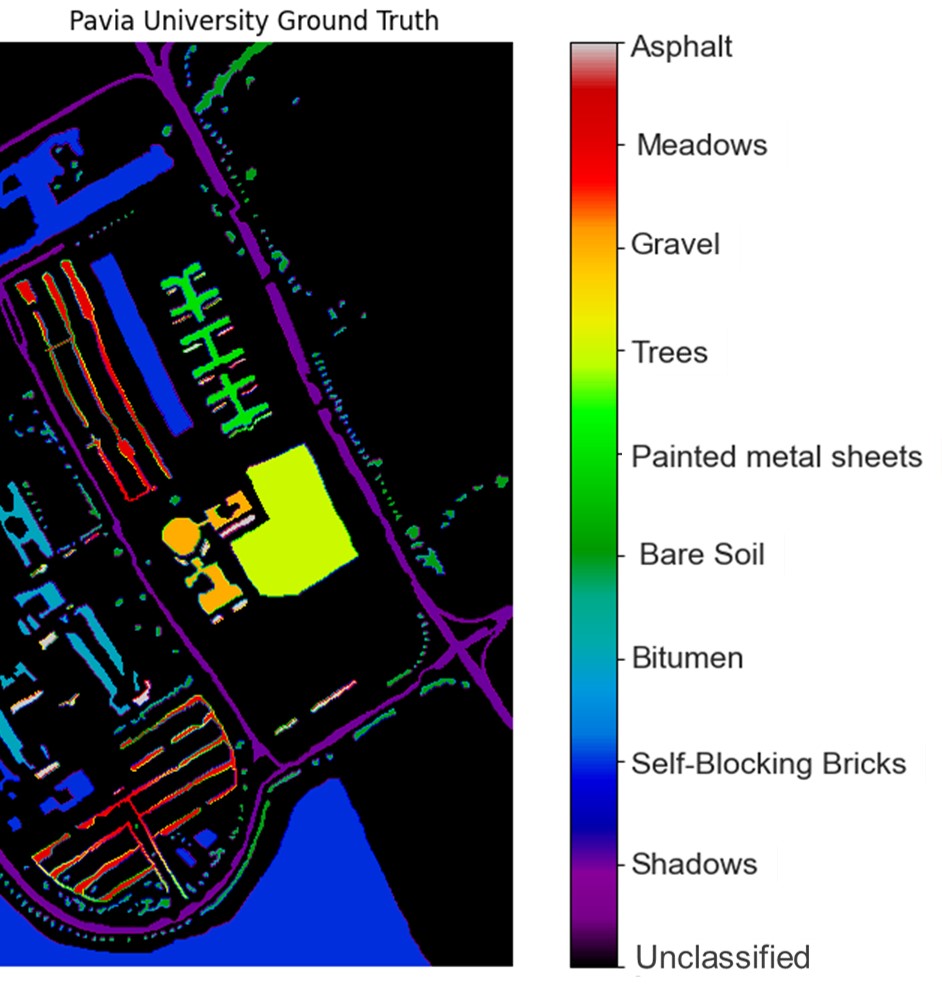}
 \vspace{-8mm}
 \caption{Ground truth image for Pavia University scene}
  %\vspace{2.0cm}
  %\centerline{(a) Result 1}\medskip
\end{minipage}
\end{figure}

\begin{table}[!ht]
\vspace{-5mm}
\caption{Ground truth classes for the Pavia University scene}
 \vspace{3mm}
\centering
\small
\begin{tabular}{|l|l|l|}
\hline
\# & Class                & Samples \\ \hline
1  & Asphalt              & 6631    \\ \hline
2  & Meadows              & 18649   \\ \hline
3  & Gravel               & 2099    \\ \hline
4  & Trees                & 3064    \\ \hline
5  & Painted metal sheets & 1345    \\ \hline
6  & Bare Soil            & 5029    \\ \hline
7  & Bitumen              & 1330    \\ \hline
8  & Self-Blocking Bricks & 3682    \\ \hline
9  & Shadows              & 947     \\ \hline
\end{tabular}
\end{table}

\section{DIMENSIONALITY REDUCTION}
\label{sec:pagestyle}

The high spectral resolution of the hyperspectral data provides advantages for the classification but also presents challenges for conventional signal-processing techniques \cite{ghamisi2017advances,ghamisiadvanced,camps1}. For such cases, dimensionality reduction techniques are utilized to map high dimensional space to lower dimensional subspace \cite{jia2013feature,camps1}. The PCA is the most commonly employed feature reduction technique in hyperspectral image classification. \par As an alternative to PCA, R-PCA, a linear dimensionality reduction technique like PCA, has been proposed for very large datasets to find approximate principal components using randomized SVD \cite{halko,Wojnowicz,firat2022}. In the experimental setup, the number of components were reduced to 20 and 30. 

\section{IMAGE CLASSIFICATION}
\label{sec:typestyle}

SVM has been the most frequently used machine learning model for classifying hyperspectral images over the past two decades. This is because SVM which is a kernel-based learning algorithm can handle high dimensional data to distinguish classes when a limited number of training data is available \cite{Gualtieri} therefore has shown great success in hyperspectral image classification \cite{melgani,pal2005,hasanlou2015svm}. The optimum parameters (C= 600 and $\gamma$= 0.5) for radial basis function kernel of SVM were determined by grid search strategy using cross validation approach. SVM classification was performed using the open-source Scikit-learn module in Python \cite{scikit}.  \par To analyze the impacts of R-PCA on image classification accuracy, another machine learning algorithm called LightGBM has been utilized in this experimental research. LightGBM is a cutting-edge ensemble learning algorithm and based on decision trees. It applies the leaf-wise (i.e., vertically) strategy while growing the trees and accelerates the training process by its two new functionalities: Gradient-based One-Side Sampling (GOSS) and Exclusive Feature Bundling (EFB) \cite{Ke,jafa}. \par In the experimental setup, the 70\% of ground truth dataset selected as training set (randomly) and the remaining parts were selected as test set.

\section{EXPERIMENTAL RESULTS AND DISCUSSION}
\label{sec:majhead}

In this experimental research, the impacts of R-PCA on hyperspectral image classification accuracy with two supervised machine learning methods (SVM and LightGBM) have been investigated and the results were compared with PCA on two hyperspectral datasets. The accuracy scores and classification maps were presented on Table 3, Table 4 and Figure 3 and Figure 4 for Indian Pines and Pavia University scenes, respectively.

%\begin{table}[htbp]
%\caption{Accuracy Scores for Indian Pines}
%\begin{center}
%\begin{tabular}{|c|c|c|}
%\hline
%\textbf{Features} & \textbf{SVM} & \textbf{LightGBM} \\ \hline
%\textbf{Original Data} & 0,9473 & 0,9639         \\ \hline
%\textbf{PCA-20}   & 0,8657 & 0,8481            \\ \hline
%\textbf{RPCA-20}  & 0,8471 & 0,3148            \\ \hline
%\textbf{PCA-30}    & 0,8803 & 0,8530            \\ \hline
%\textbf{RPCA-30}  & 0,8354 & 0,8546            \\ \hline
%\end{tabular}
%\end{center}
%\end{table}

\begin{table}[!ht]
 \vspace{-5mm}
\caption{Accuracy Scores for Indian Pines}
\centering
\begin{adjustbox}{width=0.5\textwidth}
%\small
\begin{tabular}{|cccccc|}
\hline
\multicolumn{6}{|c|}{\textbf{Indian Pines}}                                                                                                         \\ \hline
\multicolumn{1}{|c|}{\textbf{Methods}}  & \multicolumn{1}{c|}{\textbf{Original}} & \multicolumn{1}{c|}{\textbf{PCA-20}} & \multicolumn{1}{c|}{\textbf{RPCA-20}} & \multicolumn{1}{c|}{\textbf{PCA-30}} & \textbf{RPCA-30} \\ \hline
\multicolumn{1}{|c|}{\textbf{SVM}}      & \multicolumn{1}{c|}{0.9473}            & \multicolumn{1}{c|}{0.8657}          & \multicolumn{1}{c|}{0.8471}           & \multicolumn{1}{c|}{0.8803}          & 0.8354           \\ \hline
\multicolumn{1}{|c|}{\textbf{LightGBM}} & \multicolumn{1}{c|}{0.9639}            & \multicolumn{1}{c|}{0.8481}          & \multicolumn{1}{c|}{0.3148}           & \multicolumn{1}{c|}{0.8530}          & 0.8546           \\ \hline
\end{tabular}
\end{adjustbox}
\end{table}

%\begin{figure}[htbp]

%\begin{minipage}[b]{1.0\linewidth}
%\centering
 %\includegraphics[width=\textwidth]{accuracy_scores_journal_indianpines.png}
 %\vspace{-8mm}
 %\caption{Accuracy Scores for Indian Pines}
  %\vspace{2.0cm}
  %\centerline{(a) Result 1}\medskip
%\end{minipage}
%\end{figure}

\begin{figure}[htbp]

\begin{minipage}[b]{1.0\linewidth}
\centering
 \includegraphics[width=\textwidth]{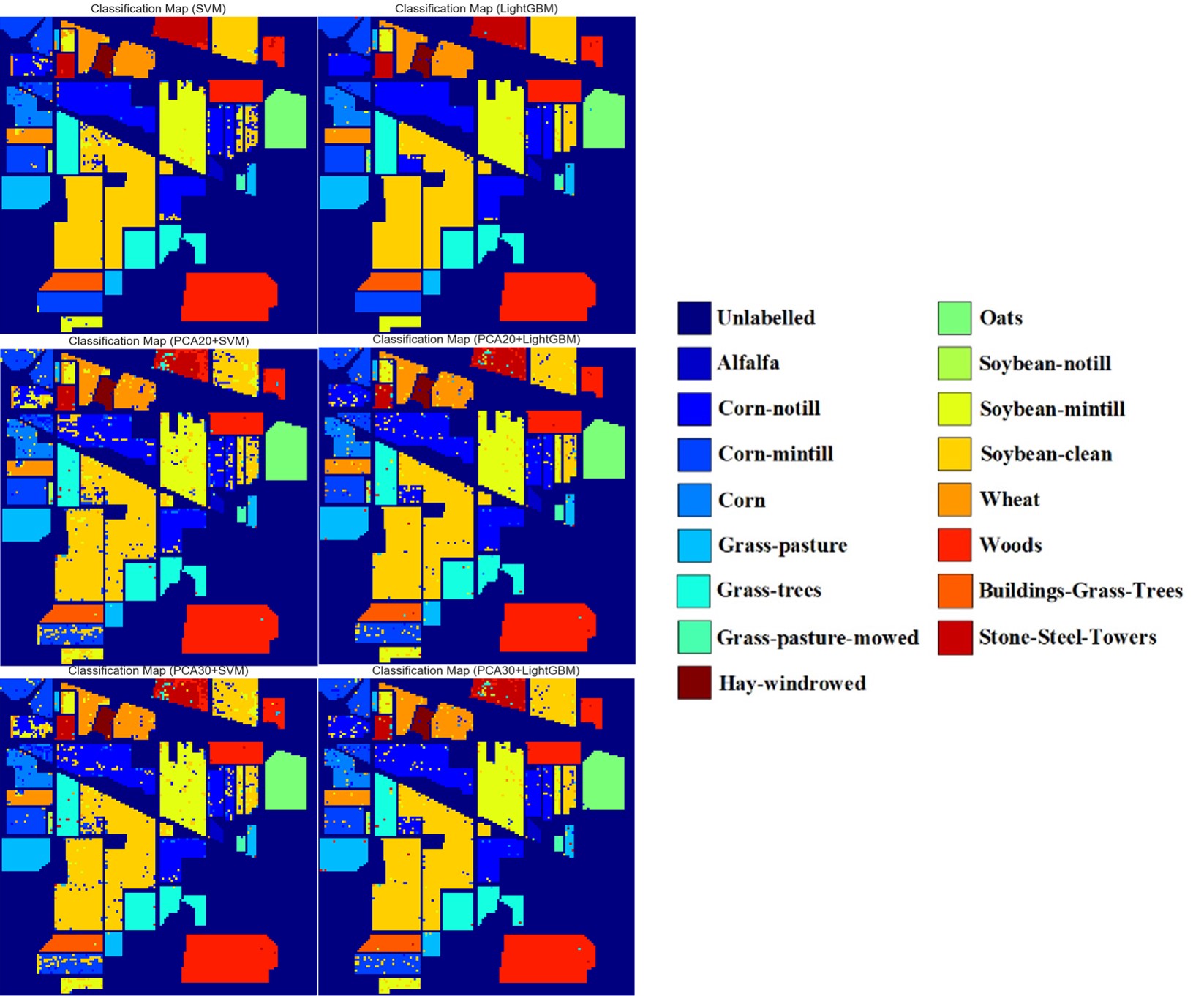}
 \vspace{-8mm}
 \caption{Classification Maps for the Indian Pines scene}
  %\vspace{2.0cm}
  %\centerline{(a) Result 1}\medskip
\end{minipage}
\end{figure}

The highest classification accuracies (overall accuracies) were received as 0.9925 and 0.9639 by LightGBM with original features for Pavia University and Indian Pines, respectively. For both datasets, the original features gives better results than the dimensional reduced features in terms of overall accuracies by SVM and LightGBM. For the Indian Pines scene, the PCA with SVM for both first 20 and 30 principal components received higher accuracies than R-PCA however R-PCA with LightGBM slightly outperformed PCA for the first 30 principal components (0.8546 vs 0.8530). \par

\begin{table}[!ht]
 \vspace{-5mm}
\caption{Accuracy Scores for Pavia University}
\centering
\begin{adjustbox}{width=0.5\textwidth}
%\small
\begin{tabular}{|cccccc|}
\hline
\multicolumn{6}{|c|}{\textbf{Pavia University}}                                                                                                         \\ \hline
\multicolumn{1}{|c|}{\textbf{Methods}}  & \multicolumn{1}{c|}{\textbf{Original}} & \multicolumn{1}{c|}{\textbf{PCA-20}} & \multicolumn{1}{c|}{\textbf{RPCA-20}} & \multicolumn{1}{c|}{\textbf{PCA-30}} & \textbf{RPCA-30} \\ \hline
\multicolumn{1}{|c|}{\textbf{SVM}}      & \multicolumn{1}{c|}{0.9760}            & \multicolumn{1}{c|}{0.9565}          & \multicolumn{1}{c|}{0.9327}           & \multicolumn{1}{c|}{0.9577}          & 0.9342           \\ \hline
\multicolumn{1}{|c|}{\textbf{LightGBM}} & \multicolumn{1}{c|}{0.9925}            & \multicolumn{1}{c|}{0.9553}          & \multicolumn{1}{c|}{0.9550}           & \multicolumn{1}{c|}{0.9540}          & 0.9539           \\ \hline
\end{tabular}
\end{adjustbox}
\end{table}

%\begin{figure}[htbp]

%\begin{minipage}[b]{1.0\linewidth}
%\centering
 %\includegraphics[width=\textwidth]{accuracy_scores_journal.png}
 %\vspace{-8mm}
 %\caption{Accuracy Scores for Indian Pines}
  %\vspace{2.0cm}
  %\centerline{(a) Result 1}\medskip
%\end{minipage}
%\end{figure}

For the Pavia University scene, the PCA with SVM for both first 20 and 30 principal components received higher accuracies than R-PCA however for the R-PCA with LightGBM, the classification accuracies are very close (equal) to those received with PCA. The statistical significance between the accuracies were evaluated by McNemar’s test (with the 95\% confidence interval) and it was found that there is no significantly difference between R-PCA and PCA for the classification of Pavia University scene with LightGBM. For the classification of the original data, the LightGBM received higher accuracy than SVM for Pavia University.

\begin{figure}[htbp]

\begin{minipage}[b]{0.9\linewidth}
\centering
 \includegraphics[width=\textwidth]{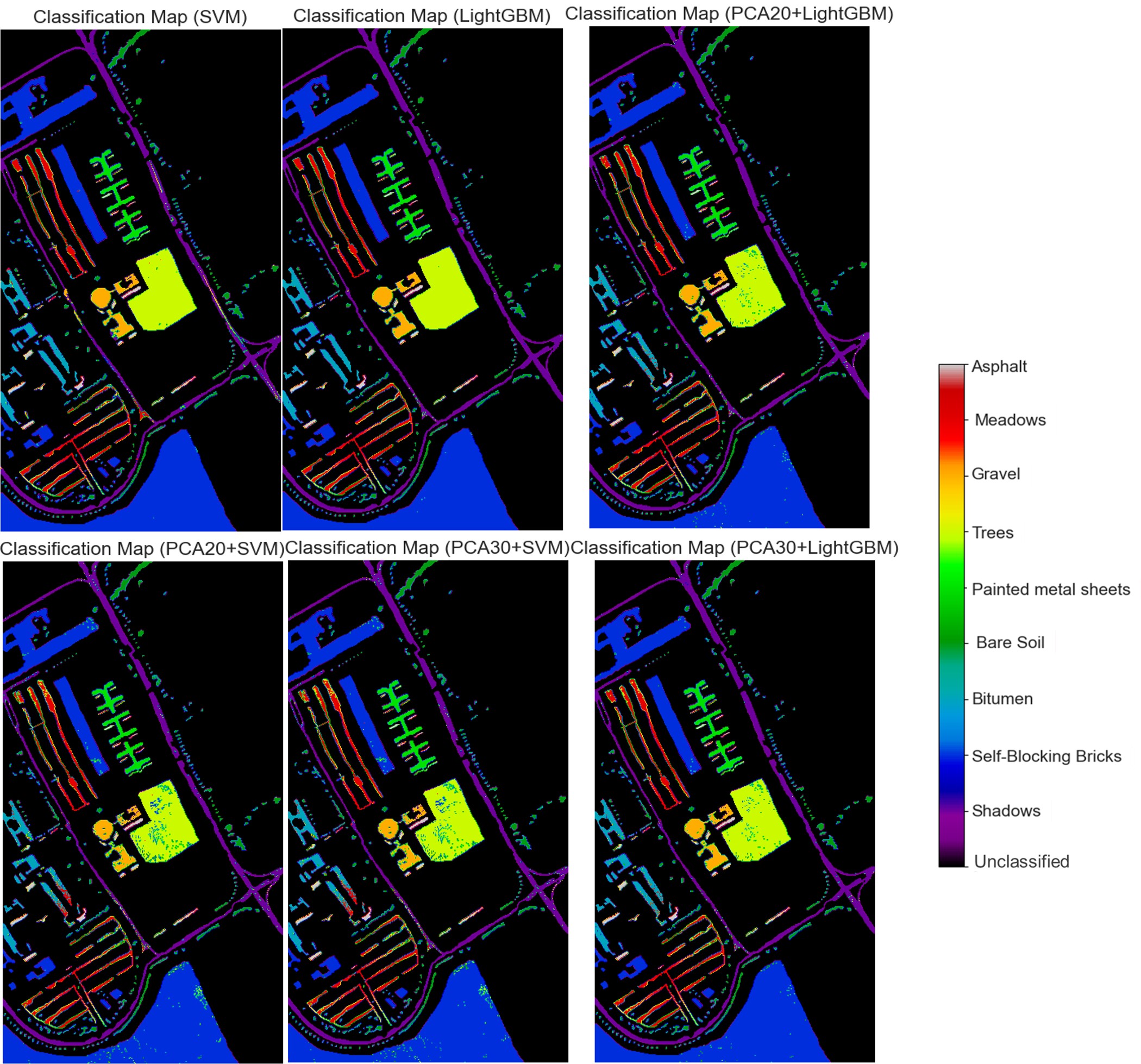}
 \vspace{-8mm}
 \caption{Classification Maps for Pavia University scene}
  \vspace{-6mm}
  %\centerline{(a) Result 1}\medskip
\end{minipage}
\end{figure}

\section{CONCLUSION}
\label{sec:conclusion}

In most of the previous studies, R-PCA was investigated along with CNN-based methods however has not been explored yet with supervised learning algorithms rather than SVM for hyperspectral image classification. With this research, the impacts of R-PCA was investigated for the classification of hyperspectral images using LightGBM and SVM and the classification performance was compared with PCA. This research has also tested the sensitivity of the kernel-based learning and ensemble learning algorithms to the same dataset for dimensionality reduction. This experimental results demonstrated that the original features yield better results than PCA and RPCA. The highest classification accuracies for the Indian Pines and Pavia University were received as 0.9639 and 0.9925 by LightGBM, respectively.

\section{ACKNOWLEDGEMENT}
\label{sec:ACKNOWLEDGEMENT}

The author would like to thank Prof. D. A. Landgrebe for providing the Indian Pines data set and Prof. P. Gamba for providing the Pavia University data set. Also, thanks to Computational Intelligence Group of the University of the Basque Country for making the datasets freely available at \cite{hyperspectral}    
% Below is an example of how to insert images. Delete the ``\vspace'' line,
% uncomment the preceding line ``\centerline...'' and replace ``imageX.ps''
% with a suitable PostScript file name.
% -------------------------------------------------------------------------

%\vfill
%\pagebreak

\bibliographystyle{IEEEbib}
\bibliography{strings,refs}

\end{document}